\documentclass[twocolumn,showpacs,amssymb,nobibnotes,nofootinbib,aps,floats,psfig,prd]{revtex4-1}
\usepackage{textcomp,amssymb,graphicx,epsf}
\usepackage{hyperref}
\usepackage{amsbsy}
\usepackage{amsmath}
\usepackage{amssymb}
\usepackage[dvips]{color}
\usepackage{graphicx,epsfig}
\usepackage{epsfig}
\usepackage{float}
\date{}
\begin{document}
\title{Tricritical points in a compact $U(1)$ lattice gauge theory at strong coupling}
\author{Asit K. De}
\author{Mugdha Sarkar}
\affiliation{Theory Division, Saha Institute of Nuclear Physics,
Kolkata, India}

\date{\today}
\begin{abstract}
{
Pure {\it compact} $U(1)$ lattice gauge theory exhibits a phase transition at gauge coupling $g \sim {\cal{O}}(1)$ separating a familiar weak coupling Coulomb phase, having free massless photons, from a strong coupling phase. However, the phase transition was found to be of first order, ruling out any nontrivial theory resulting from a continuum limit from the strong coupling side. In this work, a compact $U(1)$ lattice gauge theory is studied with addition of a dimension-two mass counterterm and a higher derivative (HD) term that ensures a unique vacuum and produces a covariant gauge-fixing term in the naive continuum limit. For a reasonably large coefficient of the HD term, now there exists a continuous transition from a regular ordered phase to a spatially modulated ordered phase. For weak gauge couplings, a continuum limit from the regular ordered phase results in a familiar theory consisting of free massless photons. For strong gauge couplings with $g\ge {\cal{O}}(1)$, this transition changes from first order to continuous as the coefficient of the HD term is increased, resulting in tricritical points which appear to be a candidate in this theory for a possible nontrivial continuum limit.    
}
\end{abstract}
\maketitle

\section{Introduction} \label{Intro}
Through a strong coupling expansion of a Wilson loop in a space-time (Euclidean) lattice, Wilson \cite{Wilson74} showed evidence for confinement in a pure {\em compact} $SU(3)$ gauge theory, marking the beginning of a new method for nonperturbative investigation of quantum field theories. Ironically, a similar calculation is equally applicable to pure $U(1)$ gauge theory that shows nontrivial properties at strong gauge coupling (the compact formulation allows self-interaction for all powers of the Abelian gauge fields on the discrete lattice). The hallmark of Wilson's approach is that gauge invariance is manifest at all stages of the calculation and gauge-fixing is not required. The theory is rigorously defined through a functional integral with a gauge-invariant (Haar) measure with group-valued gauge fields. The algebra-valued gauge fields become noncompact, smooth and dimensionful only in the continuum limit. 

Because of known physics from weak-coupling quantum electrodynamics, a $U(1)$ gauge theory, it was expected that at an intermediate gauge coupling, there would be a phase transition from the strong coupling to a familiar weak coupling phase with free massless photons in the continuum limit for the pure gauge theory. Indeed Monte Carlo simulations found this transition, which was later confirmed to be first order \cite{U1late}. Absence of a diverging correlation length meant that no quantum continuum limit could be taken in this $U(1)$ theory.

We shall now take a short detour to lattice formulations of chiral gauge theories to understand why in certain situations there is a need to control the longitudinal modes of lattice gauge fields. Fermions on the discrete lattice necessarily break chiral symmetry \cite{KS_NN, GW82}. For chiral gauge theories, obviously the gauge symmetry is then explicitly broken on the lattice. Lack of gauge invariance in the Wilson framework (without gauge-fixing) necessarily means strong coupling between the physical degrees of freedom and the longitudinal gauge degrees of freedom ({\it lgdof}s). This is explained in the following.

Because of the Haar measure, the functional integral is over all gauge configurations, including the ones related to each other by gauge transformations. As a result, after a gauge transformation, the  {\it lgdof}s become explicitly present in the action and  interact with the physical degrees of freedom. This interaction is strong because there is no gauge-fixing and any point on the gauge orbit is as likely as any other, essentially making the gauge fields very rough. 

The rough gauge problem was the main reason of failure of a full class of lattice chiral gauge theories \cite{BockDeSmit91,GolterPetcherSmit91}. These failures gave rise to the understanding that controlling the dynamics of {\it lgdof}s in these theories (in other words, gauge-fixing) is essential to avoid undesirable results. 

However, the Becchi-Rouet-Stora-Tyutin (BRST) scheme, a standard mechanism for taking care of the redundancy related to the {\it lgdof}s, cannot be used in this general nonperturbative case with compact gauge fields, because of a theorem, due to Neuberger, that proves that the partition function and the unnormalized expectation value of a gauge-invariant operator are each zero in presence of a BRST symmetry \cite{Neuberger, Testa}. This is presumably due to the cancelling contributions from a bunch of Gribov copies signalling multiple solutions of the gauge-fixing condition. 

For the general non-Abelian case, the above theorem can be evaded by employing an equivariant BRST (eBRST) formalism \cite{Schaden, GolterShamir_nAbel} where gauge-fixing is done only in the coset space, leaving, for example, an Abelian subgroup gauge-invariant. This may be taken as a viable alternate nonperturbative scheme for defining a non-Abelian gauge theory, a proposal worthy of investigation by itself. However, for a chiral gauge theory, the residual Abelian gauge symmetry needs to be fixed in an appropriate manner. Failing to do so leads to a strongly interacting sector of {\it lgdof}s which is undesirable, as explained above.    

In fact, because of the no-go theorem mentioned above, any BRST-type symmetry cannot be entertained for the Abelian theory either. A naive lattice transcription of a covariant gauge fixing term  results disastrously in a dense set of lattice Gribov copies \cite{Shamir}. To overcome this issue, Shamir  and  Golterman \cite{Shamir,GolterShamir_Abel} proposed to add, to the standard Wilson lattice gauge action for the compact $U(1)$ pure gauge case, a higher-derivative (HD) term (involving physical fields only), breaking gauge invariance explicitly. This term, as a first requirement, leads to a covariant gauge fixing term in the naive continuum limit, and, at the same time, is designed to ensure, in the weak gauge coupling limit, a unique absolute minimum for the effective potential, thus avoiding the problem of the Gribov copies and enabling weak-coupling perturbation theory (WCPT) around the unique vacuum. Counterterms are possible to construct because of the emergence of a renormalizable gauge, and are required to restore gauge symmetry.

WCPT analysis and numerical investigations performed earlier \cite{Bock_etal}, only in the weak gauge coupling region of the above compact Abelian pure gauge theory, confirmed the existence of a new continuous phase transition between a regular ordered phase and a spatially modulated ordered phase, for sufficiently large value of the coefficient of the HD term. At this phase transition, gauge symmetry is restored and the scalar fields ({\it lgdof}s) decouple, leading to the desired emergence of massless free photons only, in the continuum limit taken from the regular broken phase.

In this paper, we explore the phase diagram of the above compact $U(1)$ pure gauge theory with the HD gauge-fixing term and a suitable counterterm, {\it in the strong gauge coupling region} and present only the key findings regarding the nature of the possible continuum limits in that region. Details of our investigation will be available in \cite{DeSarkar2}.       

Our work is important from several points of view. First, for both Abelian and non-Abelian\footnote{Remember, for the non-Abelian case, the residual Abelian gauge symmetry after eBRST has to be fixed by the HD action} chiral gauge theories in the nonperturbative gauge-fixing approach, it is important to know, for a large range of the gauge couplings (including strong gauge couplings), that a correct continuum limit (with the {\it lgdof}s decoupled and massless free photons) is achievable in the pure Abelian gauge theory.\footnote{It may be mentioned here that the success of the gauge-fixing approach in lattice Abelian chiral gauge theories has been demonstrated in a number of papers \cite{ChiGT1,ChiGT2}, however, only for weak gauge couplings. The above papers involved both analytic and numerical methods in the so-called reduced $U(1)$ theory.} Second, given that the pure compact $U(1)$ gauge theory with the HD term, in the weak gauge coupling region, has produced a correct quantum continuum limit with free massless photons, it is important to explore a broader region of the coupling parameter space and ask what happens for strong gauge couplings. Obviously this question is linked with the issue of short distance behavior of $U(1)$ gauge theory and possibility of nontrivial physics. Third, HD actions are a challenge for algorithms, especially the so-called local update algorithms. The theory provides a good opportunity to evaluate aspects of different algorithms with large coefficients of the HD term. Lastly, because of the presence of tricritical points and critical endpoints in the phase diagram as we shall see, and phase transitions that restore a local symmetry, the investigated theory is interesting also from the point of view of statistical mechanics and critical phenomena. 

The paper is organized as follows. In Sec.\ref{Action}, the lattice action under investigation for the compact $U(1)$ gauge theory  is presented. The action contains a HD gauge fixing term which includes irrelevant pieces, and also a counter-term required for reviving the gauge symmetry. Section \ref{Action} also summarises the main results from previous investigations of the theory at weak gauge couplings. Section \ref{Num} presents some of the details of our numerical investigations in the parameter space of the action, including the algorithms used in our investigation. The results of our investigation are presented in Sec.\ref{Results} which identifies the tricritical point and determines the possible continuum limits. Finally, in Sec.\ref{Conclu}, we present our conclusions on the possible continuum physics from the theory investigated at strong gauge couplings.

\section{The Lattice Action} \label{Action}
The investigated lattice action is given by: 
\begin{eqnarray}
S = S_{\rm W} + S_{\rm GS} +  S_{\rm ct},
\label{S}
 \end{eqnarray} 
where,
\begin{eqnarray}
S_{\rm W} = \frac{1}{g^2}\sum_{x, \;\mu<\nu} \left( 1 - {\rm Re} \, U_{{\rm P}\mu\nu}(x) \right )
\end{eqnarray}
is the (gauge-invariant) Wilson term containing a summation over all gauge plaquettes $U_{{\rm P}\mu\nu}(x)$, the plaquette being the smallest Wilson loop on a $(\mu,\nu)$ plane.

The second term in (\ref{S}) is the Golterman-Shamir HD gauge-fixing term, given by
\begin{eqnarray}
S_{\rm GS}={\tilde{\kappa}}\left ( \sum_{xyz} \Box_{xy}(U)\, \Box_{yz}(U) - \sum_x  B_x^2 \right ),
\end{eqnarray}
with the covariant Laplacian 
\begin{eqnarray}
\Box_{xy}(U) = \sum_\mu ( \delta_{y,x+\mu} U_{x\mu} + \delta_{y,x-\mu}U^\dagger_{x-\mu,\mu} - 2 \delta_{xy} ), 
\end{eqnarray}
and,
\begin{eqnarray}
B_x = \sum_\mu ({\cal{A}}_{x-\mu,\mu} + {\cal{A}}_{x\mu})^2/4, \;
{\rm where} \; {\cal{A}}_{x\mu}={\rm Im} U_{x\mu}.
\end{eqnarray}

The third term $S_{\rm ct}$ represents possible counterterms, and we use
\begin{eqnarray}
S_{\rm ct} = -\kappa \sum_{x\,\mu} \left ( U_{x\mu} + U^\dagger_{x\, \mu}  \right )
\end{eqnarray}
which is a dimension-2 mass counterterm, as apparent from expanding the lattice gauge field $U_{x\mu} = \exp(iagA_\mu(x))$ for small lattice spacing $a$. Also possible are a host of marginal counterterms that can be treated perturbatively \cite{GolterShamir_Abel}. The dimension-2 counterterm is enough to give rise to a new universality class, as we shall see. 

The presence of the HD term ensures a unique absolute minimum for the action at $U_{x\mu}=1$, validating WCPT around $g=0$ or $\tilde{\kappa}=\infty$ and leads to the familiar covariant gauge fixing term 
\begin{eqnarray}
(1/2\xi)\int d^4x (\partial_\mu A_\mu)^2 \label{cov_gf}
\end{eqnarray}
in the naive continuum limit with $\xi = 1/(2\tilde{\kappa}g^2)$.   

\begin{figure}[t]
\includegraphics[height=4.5cm]{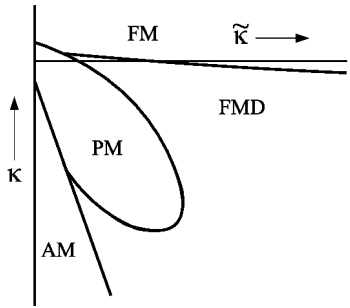}
\caption{Schematic phase diagram in the $(\tilde{\kappa},\kappa)$ plane at a given weak gauge coupling ($g<1$). 
}
\label{weakphasedia}
\end{figure}
\begin{figure}[t]
\includegraphics[width=0.99\linewidth]{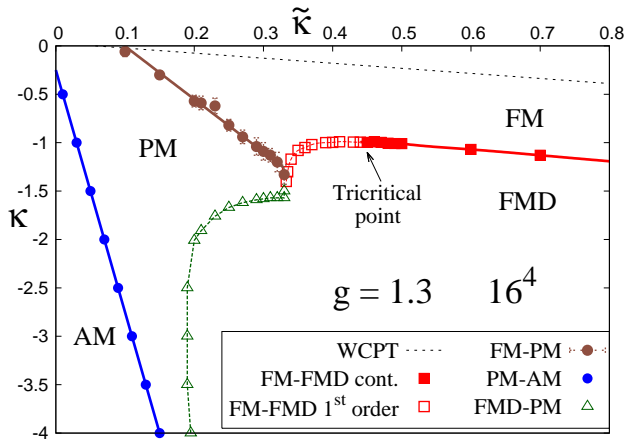}
\caption{Phase diagram in the $(\tilde{\kappa},\kappa)$ plane at gauge coupling $g=1.3$ on $16^4$ lattice. 
}
\label{strongphasedia}
\end{figure}

The action contains only physical fields (and does not include, for example, ghosts which are expected to decouple only in the continuum limit) and as such is not BRST-invariant. The relevant symmetry for the action (\ref{S}) is the gauge symmetry, and $S_{\rm ct}[U_{x\mu}]$ and $S_{{\rm GS}}[U_{x\mu}]$ are not gauge-invariant. Under a gauge transformation $U_{x\mu} \rightarrow g_xU_{x\mu}g^\dagger_{x+\mu}$, these terms pick up the {\it lgdof}s, and the theory becomes a scalar-gauge system with $S[\phi^\dagger_xU_{x\mu}\phi_{x+\mu}]$ where mass of the scalar fields $\phi_x\equiv g^\dagger_x$ with $g_x\in U(1)$ may scale in an appropriate continuous phase transition of the lattice theory. However, the goal here is not to have a gauge-Higgs theory in the continuum, it is rather to decouple the scalar fields ({\it lgdof}s) at a continuum limit.  

The two extra terms (with coefficients $\kappa$ and $\tilde{\kappa}$) ensure that in the neighborhood of the perturbative point (i.e., for small $g$ and large $\tilde{\kappa}$) the {\it lgdof}s are weakly coupled, and indeed numerical simulations confirm that the {\it lgdof}s decouple at a new phase transition separating the regular ordered phase (to be called FM in the following) from a so-called spatially modulated ordered phase (FMD) \cite{Bock_etal}.

Numerical simulations in \cite{Bock_etal}, were done at weak couplings ($g<1$) in the so-called vector picture (action (\ref{S}) where no scalars appear explicitly) and in the so-called Higgs picture (action with both scalars and gauge fields, obtained after a gauge transformation). For weak couplings, these studies confirmed a phase diagram with generic features as given in Fig.\ref{weakphasedia}.  The nomenclature of the phases in this theory has been taken as per the phases in the so-called reduced model \cite{Bock_etal}. The reduced model is obtained by putting $U_{x\mu}=1$ (corresponding to the trivial orbit) in the Higgs picture of the theory. The regular broken phase, FM (with ferromagnetic order) is characterized by a massive photon and a massive scalar, the PM (for paramagnetic) phase is the disordered (symmetric) phase having massless photons, and finally the new FMD (ferromagnetic-directional) phase is the spatially modulated ordered phase that breaks Euclidean rotational symmetry with a nonzero vector condensate ($\langle A_\mu(x)\rangle \ne 0$) (there is also an antiferromagnetic or AM phase with staggered order, not to be discussed further in this study). Photon and scalar masses scale by approaching the continuous FM-PM transition from the FM phase, leading to a continuum gauge-Higgs theory. A sufficiently large $\tilde{\kappa}$ (and small $g$) ensures a satisfactory continuum limit with only the photon mass scaling (thereby recovering gauge symmetry and decoupling the scalars) at the FM-FMD phase transition by tuning a single parameter $\kappa$ from the FM side.

Given the above that this new formulation of a compact $U(1)$ gauge theory on lattice produces a correct continuum limit for weak gauge couplings, it is certainly worthwhile to ask about the nature of a continuum limit, if at all, for strong gauge couplings and also explore the possibilities of a nontrivial theory. The strong coupling region was first explored in \cite{BasakDeSinha} with speculations of a few novel features. In this paper, a completely independent and new investigation, a more careful and precise exercise has been carried out employing new methods (see Sec.\ref{Num} below). As a result, a clear picture of the phase diagram of the theory at strong gauge couplings has emerged. In the following, we present some of the key findings, principal among them is the existence of FM-FMD transition even at strong gauge couplings, and tricritical points on this transition. 
\begin{figure}[t]
\includegraphics[width=0.99\linewidth]{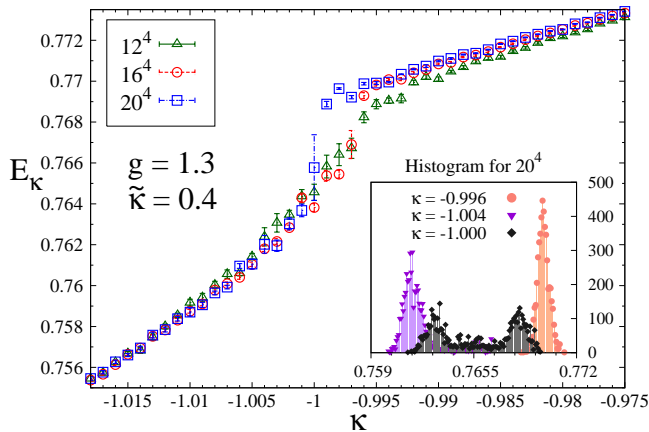}
\caption{$E_{\kappa}$ versus $\kappa$ plot on three different lattice volumes at $g=1.3$ and $\tilde{\kappa}=0.4$, showing a first order FM-FMD transition. Inset shows a histogram with a double peak structure at $\kappa=-1.000$.
}
\label{firstorder}
\end{figure}
\begin{figure}[t]
\includegraphics[width=0.99\linewidth]{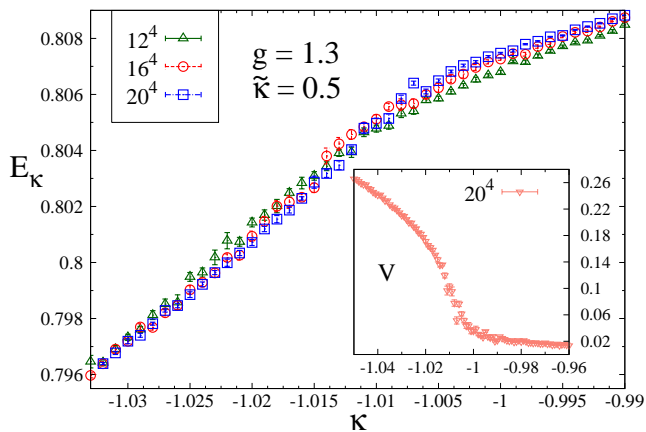}
\caption{$E_{\kappa}$ versus $\kappa$ plot on three different lattice volumes at $g=1.3$ and $\tilde{\kappa}=0.5$, showing a continuous FM-FMD transition. Inset shows $V$ versus $\kappa$ plots for the same parameters.
}
\label{secondorder}
\end{figure}

\section{Numerical simulations} \label{Num}
Multihit Metropolis, a local update algorithm that was used in all previous investigations of the theory (e.g. \cite{Bock_etal,BasakDeSinha}), was discarded for the current work because at large $g$ it produced results unstable against variation of the number of hits and also the particular order the lattice was swept. This is understandable, since with such a HD action density, spread over quite a few lattice sites, a local algorithm is bound to struggle, especially at large $\tilde{\kappa}$ (at strong gauge couplings, the FM-FMD transition is obtained at larger $\tilde{\kappa}$). In this paper, we present results of numerical simulation done with Hybrid Monte Carlo (HMC), a global algorithm, and this marks a major difference with our previous work \cite{BasakDeSinha} and produces new, reliable and numerically stable results at strong gauge couplings so that we now have a better understanding of the possible continuum limit at the FM-FMD transition at strong gauge couplings.

Numerical simulation was done at gauge couplings $g=1.0, \, 1.1, \, 1.2, \, 1.3, \, 1.5$ and also at 0.6 and 0.8 (for comparing with available results in literature) at a variety of lattice volumes $10^4,\, 12^4, \, 16^4, \, 20^4$ and $24^4$ to determine the phase diagrams in the $\kappa-\tilde{\kappa}$ plane at each fixed gauge coupling with $\kappa$-scans and $\tilde{\kappa}$-scans having intervals as fine as $\Delta\kappa = 0.001,\,\Delta\tilde{\kappa}=0.005$ around the interesting phase transition regions. Each run in the scans typically has 5000 HMC trajectories for thermalization, and 10000 - 30000 HMC trajectories for measurement. Integrated autocorrelation times were measured and taken into account for error estimates. Error bars of all data points, wherever not shown explicitly, are smaller than the symbols. Only a small fraction of our results are produced here, more details will be made available in \cite{DeSarkar2}.  

Measurements were made on lattice volume of $L^4$ (or $L^3 T$, $L\ne T$ for propagators) for the plaquette energy 
\begin{eqnarray}
E_{\rm P} = (1/(6L^4))\langle \sum_{x,\mu<\nu} {\rm Re}\, U_{{\rm P}\mu\nu}(x) \rangle, 
\end{eqnarray}
the photon mass term 
\begin{eqnarray}
E_\kappa = (1/(4L^4))\langle \sum_{x,\mu} {\rm Re}\,U_{x\mu} \rangle, 
\end{eqnarray}
the lattice version $V$ of the vector condensate $\langle A_\mu \rangle$ (expectation value of the modulus of $(1/L^4)\sum_x {\rm Im} \,U_{x\mu}$ averaged over all the directions), the photon propagator and also the chiral condensate with quenched Kogut-Susskind (KS) fermions. The vector condensate $V$ is the order parameter for the FM-FMD transition. 

\section{Results} \label{Results}
In Fig.\ref{strongphasedia} we show the phase diagram at a fixed strong coupling $g=1.3$ in the $\kappa-\tilde{\kappa}$ plane (with a certain criterion to determine the transitions on a finite lattice). Similar phase diagrams were also obtained for $g=1.1,\, 1.2, \, 1.3, \, 1.5$. Contrast the above with that of Fig.\ref{weakphasedia}. The FM-FMD phase transition in Fig.\ref{weakphasedia} for $g<1$ is entirely a continuous transition \cite{Bock_etal,BasakDeSinha}, while for strong $g$ (Fig.\ref{strongphasedia}) there is a tricritical point separating a first order FM-FMD transition from a continuous transition. Figure \ref{strongphasedia} shows the tricritical point for $g=1.3$ at $\kappa=-0.99\pm0.01$ and $\tilde{\kappa}=0.45\pm0.02$ for the lattice size $16^4$. The location changes slightly with lattice volumes bigger than $10^4$. The PM-FMD transition is found to be strongly first order, and ends at a critical endpoint  \cite{Somen} where the continuous FM-PM transition terminates at first order transitions. The first order FM-FMD transition weakens gradually as $\tilde{\kappa}$ is increased till the tricritical point where it becomes continuous.

The location of the triciritical point in the $\kappa-\tilde{\kappa}$ plane shifts to more negative $\kappa$ and also to larger $\tilde{\kappa}$ with increasing $g$. It appears from our simulation that, at a particular gauge coupling $g^*$ between $g=1.0$ and 1.1, the tricritical point tends to approach the critical endpoint. For $g<g^*$, the FM-FMD transition is fully continuous.

The rest of the plots in this paper are all at $g=1.3$.

Figures \ref{firstorder} and \ref{secondorder} plot $E_\kappa$, an observable similar to the entropy for the $\kappa$ scan.   With increasing volume, Fig.\ref{firstorder} shows a more distinct gap at $\kappa_{\rm FM-FMD} \sim -1.00$  for $\tilde{\kappa}=0.4$, while Fig.\ref{secondorder} shows no discontinuity at $\kappa_{\rm FM-FMD} \sim -1.01$ for $\tilde{\kappa}=0.5$. The inset of Fig.\ref{firstorder} shows a double peaked histogram at the critical $\kappa$, confirming the transition to be first order. The inset of Fig.\ref{secondorder} shows the corresponding $V$ versus $\kappa$ plot illustrating the FM-FMD transition at $\kappa \sim -1.01$ for $\tilde{\kappa}=0.5$.  
\begin{figure}[t]
\includegraphics[width=0.99\linewidth]{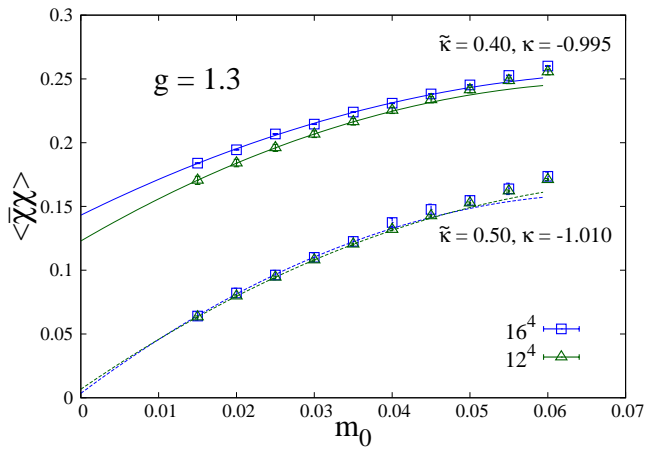}
\caption{Quenched chiral condensate in the FM phase near FM-FMD transition: nonzero at $\tilde{\kappa}=0.4$ (first order FM-FMD), and consistent with zero at $\tilde{\kappa}=0.5$ (continuous FM-FMD) at $g=1.3$.     
}
\label{chiralcond}
\end{figure}
\begin{figure}[t]
\includegraphics[width=0.99\linewidth]{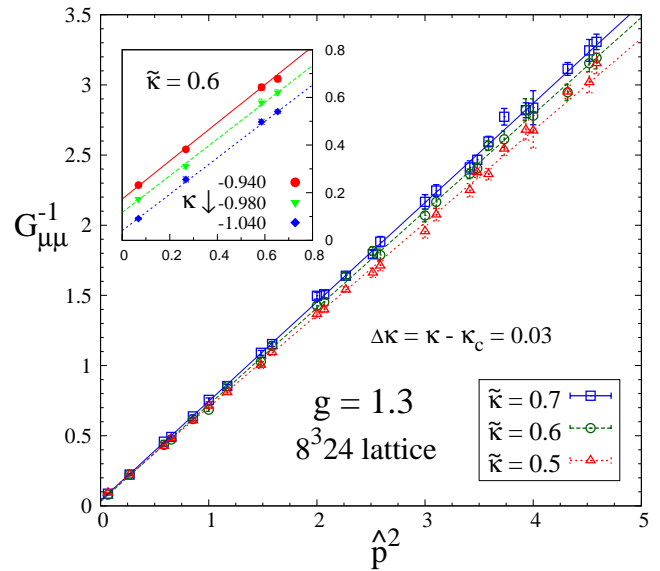}
\caption{Inverse photon propagators in the FM phase near FM-FMD transition at $g=1.3$ and at three values of $\tilde{\kappa}$ for which the transition is continuous. Inset shows scaling of the photon mass as FM-FMD transition is approached from FM side.   
}
\label{photonprop}
\end{figure}

To understand the properties of the FM phase around the tricritical point, the region was further probed with quenched KS fermions. Figure \ref{chiralcond} shows, for two lattice volumes, quenched chiral condensates in the FM phase near the FM-FMD transition at $\tilde{\kappa} = 0.4$ (where the transition is first order) and at  $\tilde{\kappa} = 0.5$ (where the transition is continuous). Noisy estimator method  was employed using 4 noise vectors with conjugate gradient inverter (Bi-CGStab, a more modern inverter, was also tried without any gain). Extrapolation to zero bare fermion mass $m_0$ was done with a phenomenological polynomial ansatz (keeping up to quadratic terms with five lowest masses fitted) and shows a condensate consistent with zero on the continuous side of the FM-FMD transition while clearly there is a nonzero condensate where the transition is first order. There is a hint of nonzero chiral condensate as the tricritical point is approached from the FM side. However, confirmation on larger volumes is required.     
    
The inverse of photon propagator (2-point correlator of ${\rm Im}\, U_{x\mu}$) in momentum space was also measured and is plotted against the square of lattice momentum $\hat{p}^2$ (discrete on a finite box) in Fig.\ref{photonprop} for the continuous part of the FM-FMD transition, staying in the FM phase.\footnote{It may be mentioned here that the three data lines in the main part of Fig.\ref{photonprop} at three $\tilde{\kappa}$ are each at a value of $\kappa$ which is at a fixed small distance away ($\Delta \kappa = \kappa - \kappa_{\rm FM-FMD}= 0.03$) from the FM-FMD transition.} Inset shows a gradually vanishing photon mass ($y$-intercept), as $\kappa$ approaches $\kappa_{\rm FM-FMD}$ ($\sim -1.07$) for the given fixed $\tilde{\kappa}$ (0.6), suggesting an expected scaling of the photon mass at the transition and recovery of gauge symmetry. The slope of the fitted straight lines, in the main figure, suggests a field renormalization constant $Z$ that is not unity. However, the figure shows that the slope increases with increasing $\tilde{\kappa}$. It seems reasonable to expect the slope to approach unity at large $\tilde{\kappa}$, consistent with WCPT at $g=0$ or $\tilde{\kappa}=\infty$. In addition, the continuous FM-FMD transition line at all strong $g$ is found in our simulations to be below but roughly parallel to the transition (the dotted line in Fig.\ref{strongphasedia}), obtained from 1-loop WCPT in \cite{Bock_etal}. Of course, the gap between them decreases as the coupling gets smaller.  

\section{Conclusion} \label{Conclu}
The phase diagram of the compact $U(1)$ pure gauge theory in the nonperturbative gauge-fixing approach including the HD term and a mass counterterm has turned out to be somewhat more complex in the strong gauge coupling region ($g>1$). Numerical simulation is also more difficult in this region, and local update algorithms struggle, forcing us to use global update algorithms with some care. However, after the algorithmic issues were sorted out, a clear picture of the possible continuum limits has emerged which is very relevant for all the major important issues discussed in Sec.\ref{Intro}, e.g., both Abelian and non-Abelian lattice chiral gauge theories and short distance behavior of $U(1)$ gauge theories.

Existence of the FM-FMD transition at strong gauge couplings is confirmed. The continuous part of this transition, away from the tricritical point, appears to produce familiar physics with free massless photons (and {\it lgdof}s decoupled) and zero chiral condensate. 

The possibility for a nontrivial continuum limit in this pure compact $U(1)$ lattice gauge theory at strong gauge couplings rests on the tricritical points with a new universality class. 

Details of our study including results at other gauge couplings that help develop the overall picture in the strong gauge coupling region, will appear in \cite{DeSarkar2}. Based on evidences so far, the gauge-fixing scheme appears as a valid method to define a gauge theory nonperturbatively. For the non-Abelian case, this involves eBRST (gauge-fixing the coset) and preparations are underway to study it as well.

\section*{Acknowledgement}
The work is fully supported by the Department of Atomic Energy (DAE), Government of India. The authors acknowledge the computing facilities of the Theory Division, Saha Institute of Nuclear Physics under DAE.

\end{document}